**Title**
Gravity Prior and Temporal Horizon Shape Interceptive Behavior under Active Inference

*Marta Russo[1], Antonella Maselli[2], Federico Maggiore[1], Giovanni Pezzulo[1]*

[1]Institute of Cognitive Sciences and Technologies, National Research Council (CNR), Via Giandomenico Romagnosi 18A, 00196, Rome, Italy
[2]Department of Biomedical and Dental Sciences and Morphofunctional Imaging, University of Messina, Via Consolare Valeria, 1, 98124 Messina, Italy

**Abstract**

Accurate interception of moving objects, such as catching a ball, requires the nervous system to overcome sensory delays, noise, and environmental dynamics. One key challenge is predicting future object motion in the presence of sensory uncertainty and inherent neural processing latencies. Theoretical frameworks such as internal models and optimal control have emphasized the role of predictive mechanisms in motor behavior. Active Inference extends these ideas by positing that perception and action arise from minimizing variational free energy under a generative model of the world.

In this study, we investigate how different predictive strategies and the inclusion of environmental dynamics, specifically an internal model of gravity, influence interceptive control within an Active Inference agent. We simulate a simplified ball-catching task in which the agent moves a cursor horizontally to intercept a parabolically falling object. Four strategies are compared: short temporal horizon prediction of the next position or long horizon estimation of the interception point, each with or without a gravity prior. Performance is evaluated across diverse initial conditions using spatial and temporal error, action magnitude, and movement corrections.

All strategies produce successful interception behavior, but those that incorporate gravity and longer temporal horizons outperform others. Including a gravity prior significantly improves spatial and temporal accuracy. Predicting the future interception point yields lower action values and smoother trajectories compared to short-horizon prediction. Furthermore, model robustness analyses demonstrate that Active Inference agents maintain performance across moderate levels of sensory noise.

These findings suggest that internal models of physical dynamics and extended predictive horizons can enhance interceptive control, providing a unified computational account of how the brain may integrate sensory uncertainty, physical expectations, and motor planning in real time.

**Corresponding author:**
Marta Russo
marta.russo@istc.cnr.it
Institute of Cognitive Sciences and Technologies, National Research Council (CNR), Via Giandomenico Romagnosi 18A, 00196, Rome, Italy

## Introduction

Catching or intercepting a flying object is a seemingly simple task that belies the complexity of the underlying neural computations. To successfully catch a ball or avoid an incoming object, the brain must overcome the inherent delays and noise in sensory processing and motor execution. Visual information about an object's position and velocity reaches cortical areas with a delay of approximately 100 ms [1,2], during which a fast-moving object can traverse a significant distance. Additionally, the generation of an appropriate motor command introduces further latency. In the field of motor control, it is still debated how the brain supports accurate catching and interception in the face of these temporal and sensory constraints [3,4].

Several theoretical frameworks have proposed strategies by which the central nervous system may cope with these challenges to achieve the exquisite abilities humans display in motor control [5–7]. Depending on the assumed temporal horizon of action and on the dynamics attributed to the environment, different models of interception have been developed. These range from the tau model [8,9], according to which a stereotyped motor command is automatically triggered when the time-to-contact reaches a critical threshold [10], to approaches such as the goal-keeper model, which propose that interception is achieved by continuously coupling the effector with the moving object [11]. Subsequent extensions of these models have incorporated both the speed and direction of object motion to refine interception strategies [12–15].

A more recent and widely accepted perspective suggests that interception behavior likely relies on a combination of online sensory adjustments and future-oriented extrapolation of object trajectories [4]. This perspective is based on the idea that the brain employs internal models to control movement and to predict the dynamics of external objects and compensate for sensory delays.

An efficient interception strategy should be adapted to the dynamics of the environment in which the agent operates. In everyday interactions, gravitational acceleration represents a fundamental and invariant property of object motion. To cope with such environment, several studies suggest that the brain encodes an internal model of gravity, allowing it to account for the effect of gravitational acceleration in predicting the object's motion [16–20]. This hypothesis is supported by behavioral and neurophysiological evidence showing that humans predict motion consistent with Earth's gravity (1g) more accurately than motion under altered gravitational conditions [21–24].

Traditional computational approaches to modeling predictive mechanisms rely on Bayesian estimators that integrate noisy sensory inputs with predictions from internal models [25–31]. These estimators provide a normative framework for (optimal) state estimation, which in turn can be used to steer (optimal) motor control [32–34]. Recent developments in computational neuroscience propose an extension of the Bayesian formulation—Active Inference—wherein both state estimation and motor control emerge from a common imperative: the minimization of variational free energy under a generative model of the environment [35–42].

Despite substantial empirical and theoretical advances [43–45], there is still no consensus on a unified computational framework that jointly accounts for prediction, efficient control, and the role of physical constraints such as gravity in interceptive actions. Taking object interception as a paradigmatic example of a challenging sensorimotor task, we address the following questions: how does an agent control its actions during interception? To what extent are environmental dynamics, and in particular gravity, incorporated into predictive mechanisms? And what temporal horizon is most effective for estimating the location at which interception will occur?

From a computational perspective, these questions translate into assessing the advantages and disadvantages of different predictive horizons and of incorporating gravity as an invariant in behavioral control.

To this end, the present study aims to test Active Inference models of interceptive behavior. Specifically, we investigate the computational roles of an internal model of gravity and of short versus long temporal prediction horizons in steering accurate object interception. We present Active Inference simulations of ball interception, comparing strategies that either include or omit a prior expectation of gravitational acceleration. Furthermore, we explore structural differences in the temporal extent of prediction by contrasting *Interceptive Location* strategies, which estimate the endpoint of an object's trajectory, with *Next Location* strategies, which predict object position over shorter temporal horizons.

This approach offers a novel computational account of interceptive behavior, linking the concept of internal models to the principles of Active Inference, and providing insights into how the brain may integrate sensory uncertainty, physical priors, and motor planning in real time.

**Methods**
This study investigates different interceptive strategies for intercepting a falling ball using a simulated active inference agent. The approach involves modeling the agent's interaction with the environment and computing error metrics to evaluate its performance. The simulation assesses four distinct interceptive strategies. The first two strategies predict the location where the interception will take place, with (*Interceptive Location With Gravity*) or without (*Interceptive Location No Gravity*) an internal model of gravity. The last two strategies predict with a shorter temporal horizon the location where the agent will aim to, with (*Next Location With Gravity*) or without (*Next Location No Gravity*) an internal model of gravity.

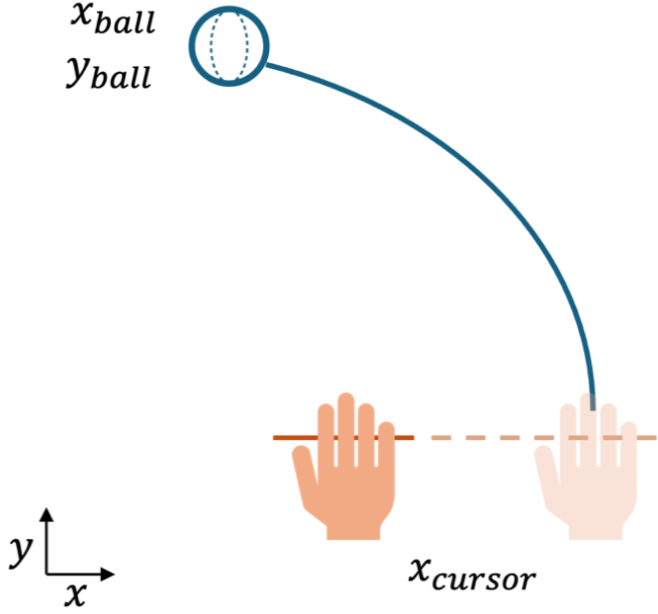

**Figure 1**: Illustration of the ball interception task. The agent attempts to intercept a falling ball by adjusting its position along the x-axis. The ball follows a parabolic trajectory under gravitational acceleration. The dashed line represents the agent line of motion, while the blue solid line shows the full ball's trajectory.

*Simulation Setup*

In our simulations, an active inference agent moves a cursor on the horizontal dimension to catch a ball projected from the top of the screen with both horizontal and vertical velocity. The scenario is in 2-D, as illustrated in Figure 1, similar to a pong arcade game.

The simulation models a ball moving under gravitational acceleration $g = -9.81 \frac{m}{s^2}$, with an initial horizontal velocity $v_{b0} = 5 m/s$ and zero vertical velocity, starting at a location 5m higher than the cursor movement line. The agent was starting at an initial x-coordinate $x_{c0} = 0m$, same as the ball horizontal coordinate. The dynamics of ball and cursor are represented by state-space equations. The agent receives noisy sensory inputs and updates its belief states using standard active inference routines, see below and [46].

The system's kinematics is discretized with a time step of $dt = 0.001 \, s$, and the total simulation duration is $1s$.

Simulation and analysis were implemented using customized software written in MATLAB R2023a. Random seeds were fixed using *rng(6)* for reproducibility.

*Active inference agent: generative process and generative model*

As in previous publications we implemented an active inference agent. In the active inference framework, the "real" physical process that describes the physical environment (here the states of the ball and of the agent and their dynamics) is usually called *generative process*, whereas the agent's internal model of the physical process is called *generative model* [47–49].

The *generative process* used in this simulation is described by the following equations:

$$x = \begin{bmatrix} x_{ball} \\ v_{xball} \\ y_{ball} \\ v_{yball} \\ x_{cursor} \\ v_{cursor} \end{bmatrix} \quad (1)$$

$$\dot{x} = f_x(x) = \begin{bmatrix} \dot{x}_{ball} \\ 0 \\ \dot{y}_{ball} \\ g \\ \dot{x}_{cursor} \\ a - \frac{c}{m} v_{cursor} \end{bmatrix} \quad (2)$$

$$\rho = g_x(x) + \sigma = \begin{bmatrix} x_{ball} \\ v_{xball} \\ y_{ball} \\ v_{yball} \\ x_{cursor} \\ v_{cursor} \end{bmatrix} + \begin{bmatrix} \sigma_{xball} \\ \sigma_{vxball} \\ \sigma_{yball} \\ \sigma_{vyball} \\ \sigma_{cursor} \\ \sigma_{vcursor} \end{bmatrix} \quad (3)$$

Equation 1 describes the state of the system, defining the state vector $x$ in terms of the ball's and the cursor's instantaneous positions and velocities. Equation 2 describes the system's dynamics $\dot{x}$. We describe the system in Cartesian coordinates approximated to the first order both for the ball (both coordinates) and the cursor, and we assume it evolves in time under the effect of external forces (i.e. gravity) and internal forces (i.e. the agent's action $a$). In particular, we assume that the state of the cursor is affected by the agent's actions ($a$ in eq. 2). The action onset time is set at a quarter of the simulation time, micking the latency of the central nervous systems for processing the sensory information and generates a motor response.

Equation 3 describes the sensory state vector $\rho$ in which we considered only visual feedback; for simplicity, we omit the proprioceptive feedback from the agent's effector, since in our setup visual and proprioceptive information are largely redundant. We assume that the sensory inputs $\rho$ are a noisy measure of position and velocity of both ball and cursor, i.e., $g_x = I$. This choice has been motived by several studies which assessed that acceleration is not key to time interceptive actions [50–52]. The sensory variance is set to $\sigma_{xball} = \sigma_{yball} = \sigma_{xcursor} = 0.01m$ and $\sigma_{vxball} = \sigma_{vyball} = \sigma_{vcursor} = 0.1m$.

The agents' internal representation of the world – the *generative model* – is described by equations 4 and 5.

$$\boldsymbol{\mu} = \begin{bmatrix} \mu_{x_{ball}} \\ \mu_{v_x ball} \\ \mu_{y_{ball}} \\ \mu_{v_y ball} \\ \mu_{x_{cursor}} \\ \mu_{v_{cursor}} \end{bmatrix} \quad (4)$$

$$\dot{\boldsymbol{\mu}} = \boldsymbol{f}_\mu(\boldsymbol{\mu}, \mu_{x_T}) = \begin{bmatrix} \dot{\mu}_{x_{ball}} \\ 0 \\ \dot{\mu}_{y_{ball}} \\ \dot{\mu}_{v_y ball} \\ \dot{\mu}_{x_{cursor}} \\ -\frac{k}{m}\left(\mu_{x_{cursor}} - \mu_{x_T}\right) - \frac{c}{m}\mu_{v_{cursor}} \end{bmatrix} \quad (5)$$

We assumed that the internal state vector $\boldsymbol{\mu}$ and its dynamics $\dot{\boldsymbol{\mu}}$ resemble the dynamics of the environment. The cursor is modeled as a Mass-Spring-Damper (MSD), with a damping factor $-\frac{c}{m}\mu_{v_{cursor}}$ to account for resistance components characterizing physical systems and to ensure that movements velocity does not increase indefinitely. The intended target is indicated as a point attractor $\mu_{x_T}$.

The MSD had been set in order to have the system critically damped, i.e. to avoid possible divergent behavior. Therefore, for a spring constant $k = 15$ and a mass of $.05\ kg$ the damping coefficient is $c = \sqrt{4km}$.

*Implementation of Predictive Control Strategies*

We realized four active inference agents, which only differ in how the point attractor $\mu_{x_T}$ is calculated. To design the four agents, we manipulated two factors. First, we manipulated the way the agent updated its own motion; for this, two agents controlled the state of the cursor following the ball, which is predicting the ball location after 100ms and setting as the attractor its estimated location along the x-axis. The other two agents controlled the cursor always aiming it at the estimated location in which the ball will intercept the y-axis (location of possible interception). Second, we manipulated the role of gravity in the estimate of ball trajectory; for this, two agents had a gravity prior and the two other agents did not. These manipulations gave rise to 4 possible agents (see Table 1):

*'Next Location No Gravity'*: Assumes constant-velocity ball motion and directs the cursor toward an attractor corresponding to the ball's estimated position over a short temporal horizon (100 ms).

*'Interceptive Location No Gravity'*: Computes the interception point under a constant-velocity assumption, with the cursor attracted to the estimated location at which interception is expected to occur.

*'Next Location With Gravity'*: Incorporates a gravity prior in the ball's motion and estimates the cursor attractor over a short temporal horizon.

*'Interceptive Location With Gravity'*: Estimates the cursor attractor at the predicted interception point, assuming gravitational acceleration governs the ball's motion.

For each strategy the attractor position $\mu_{xT}$ is determined at every time step $i$, updating it dynamically as new sensory information is received. Based on the different assumptions of each strategy, the point attractor $\mu_{xT}$ will be determined by a set of equations as described in Table1.

It must be noticed that the system, as defined in Equations 1 and 2, is deterministic and it allows for a unique possible interception point between the ball trajectory and all possible cursors trajectories. Nevertheless, the agent does not have direct access to this information, and it needs to set a strategy to intersect the ball based on its current belief.

|  | **No Gravity** | **With Gravity** |
|---|---|---|
| Next Location | $\Delta t = 100ms$<br>$\mu_{xT}(i+1) = x(i) + v_{xball}(i)\Delta t$ | $\Delta y = 50mm$<br>$\Delta t = \frac{1}{g}\left(-v_{yball}(i) - \sqrt{v_{yball}(i)^2 - 2g\Delta y_{ball}}\right)$<br>$\mu_{xT}(i+1) = x(i) + v_{xball}(i)\Delta t$ |
| Interceptive Location | $T_{end} = -\frac{y_{ball}(i)}{v_{yball}(i)}$<br>$\mu_{xT} = x(i) + v_{xball}(i)T_{end}$ | $T_{end} = \frac{1}{g}\left(-v_{yball}(i) - \sqrt{v_{yball}(i)^2 - 2gy_{ball}(i)}\right)$<br>$\mu_{xT} = x(i) + v_{xball}(i)T_{end}$ |

**Table 1.** Attractor position definitions. Depending on the strategy, the target position at which the system aims, $\mu_{xT}$, is defined with different assumptions. No Gravity and With Gravity columns respectively consider gravity as 0 or $g = -9.81 \frac{m}{s^2}$. The rows indicate if the system follows the ball (*Next Location*) or aims at the estimated location of the unique available interception point (*Interceptive Location*).

*System update*

At each time step, the inferred state of the ball and the cursor $\boldsymbol{\mu}$, and the corresponding ensuing action $a$ are updated following a gradient descent rule that aims at minimizing the Variational Free Energy of the system, as indicated in equations 6 and 7.

$$\boldsymbol{\mu}(i+1) = \boldsymbol{\mu}(i) + dt\left[\dot{\boldsymbol{\mu}}(i) - \boldsymbol{k}_\mu \frac{\partial F}{\partial \boldsymbol{\mu}}(\boldsymbol{\mu}(i), \boldsymbol{\rho}(i))\right] \qquad (6)$$

$$a(i+1) = a(i) - dt\left[\boldsymbol{k}_a \frac{\partial F}{\partial a}(\boldsymbol{\mu}(i), \boldsymbol{\rho}(i))\right] \qquad (7)$$

The gains $k_\mu$ and $k_a$ modulate the rate of gradient descent, with a defined gain $k_a = 50$ for the action update. The Variational Free Energy $F$ is defined as in eq. 8, where $\varepsilon_\mu$ and $\varepsilon_\rho$ are the model prediction errors and the sensory prediction errors, respectively.

$$F = \tfrac{1}{2}\sum \left(\frac{\varepsilon_\rho^2}{\sigma_\rho} + \frac{\varepsilon_\mu^2}{\sigma_\mu}\right) + C \tag{8}$$

$$\varepsilon_\rho = g_x(x) - \mu \tag{9}$$
$$\varepsilon_\mu = \dot{\mu} - f_x(x) \tag{10}$$

For each strategy implementation, the gains $k_\mu$ relative to the cursor update were optimized. To avoid oscillations and divergent behavior of the harmonic oscillator, the integral of the difference of the position of the cursor and of the ball in the x -coordinate has been used as a cost to optimize the gains. A custom script in MATLAB defined the cost function and used the *fmincon* function with the Sequential Quadratic Programming algorithm to find the best set of gains for each strategy.

*Error Analysis*

To evaluate each agent's performance four error metrics have been calculated. *Action Magnitude* at the instant when the ball crosses the cursor movement's line (i.e. when $y_{ball} = 0$ for the first time) is the force internally generated to act on the external environment exerted by the agent. *Spatial Error* is the absolute difference between the ball's position and the position of the cursor at the time instant when the ball reaches $y_{ball} = 0$. *Temporal Error* is the absolute time difference between the time at which the ball reaches $y_{ball} = 0$ and the time the cursor got closer (i.e. minimum distance) to the ball trajectory. Additionally, the number of velocity sign changes in the cursor's movement were counted as *Inversions*. Only changes in sign of a magnitude at least half of the peak velocity of the cursor were considered.

3D error manifolds were calculated by varying $x_{c0}$ (initial x-coordinate of the agent $x_{cursor}(t=0) = x_{c0}$) and $v_{b0}$ (initial ball horizontal velocity $v_{xball}(t=0) = v_{b0}$) across predefined ranges, which were $x_{c0} = [\,0, 0.5, 1, 1.5, 2, 2.5, 3, 3.5, 4, 4.5, 5, 5.5, 6, 6.5]$ and $v_{b0} = [1.5, 2, 2.5, 3, 3.5, 4, 4.5, 5, 5.5, 6, 6.5, 7, 7.5, 8]$ . The errors are computed for each combination of $x_{c0}$ and $v_{b0}$ values and visualized using surface plots. Note that the initial velocity vector $v_{xball}$ has always the vertical component null, so $v_{xball}$ can be specified by the scalar value of the horizontal velocity.

To assess the robustness of the model to sensory uncertainty, we systematically manipulated the level of Gaussian noise added to the sensory channels. More details can be found in the Supplementary Materials.

*Statistical Analysis*

The four metrics chosen to evaluate the interceptive strategies have been compared with a two-ways ANOVA which had as independent variables *gravity* (i.e. if the strategy took into account gravity in the estimation) and *temporal horizon* (i.e. if the strategy predicted the *next location* or the *interceptive location* of the cursor path). The entire ranges of initial cursor locations $x_{c0}$ and initial ball horizontal velocities $v_{b0}$ used for the analysis of the errors have been considered for the statistical tests.

**Results**

In this section we show results from the four strategies adopted by the agent to move the cursor towards the interception point. Each simulation follows the system's dynamics for 1 second, which is 42 ms longer that the time needed for the ball to intercept the y-axis. Initial conditions are the same for each strategy, with $x_{c0} = 0m$ and $v_{b0} = 5\frac{m}{s}$, and for each strategy we display all the relevant variables for one run of the simulation with the optimized gains.
Overall, every strategy guided the cursor close to the ball's interception point, yet the different implementations produced distinct system responses to the same stimuli.

*Next Location No Gravity Strategy*
In this simulation the agent moves the cursor to intercept a falling ball. The agent's prediction excludes a gravity prior (although the ball is subject to it) and forecasts the target's position at a short time interval ahead (100ms). In the example described in the following, the cursor starts at the same horizontal coordinate as the ball.
The agent's generative model reproduces the ball's motion very accurately (Figure 2A). Figure 2B plots the cursor's internal estimate and the executed movement over time: the estimated position rises steadily toward the interception point, while the action produces an initial burst (as soon as it is unlocked at time 0.25s) and then tracks the estimate. Importantly, the action never falls to zero. Figures 2C and 2D depict ball and cursor trajectories in the x–y plane for the internal estimate and the real environment, respectively, with darker colours marking later times. The cursor moves smoothly toward the ball and intercepts it.

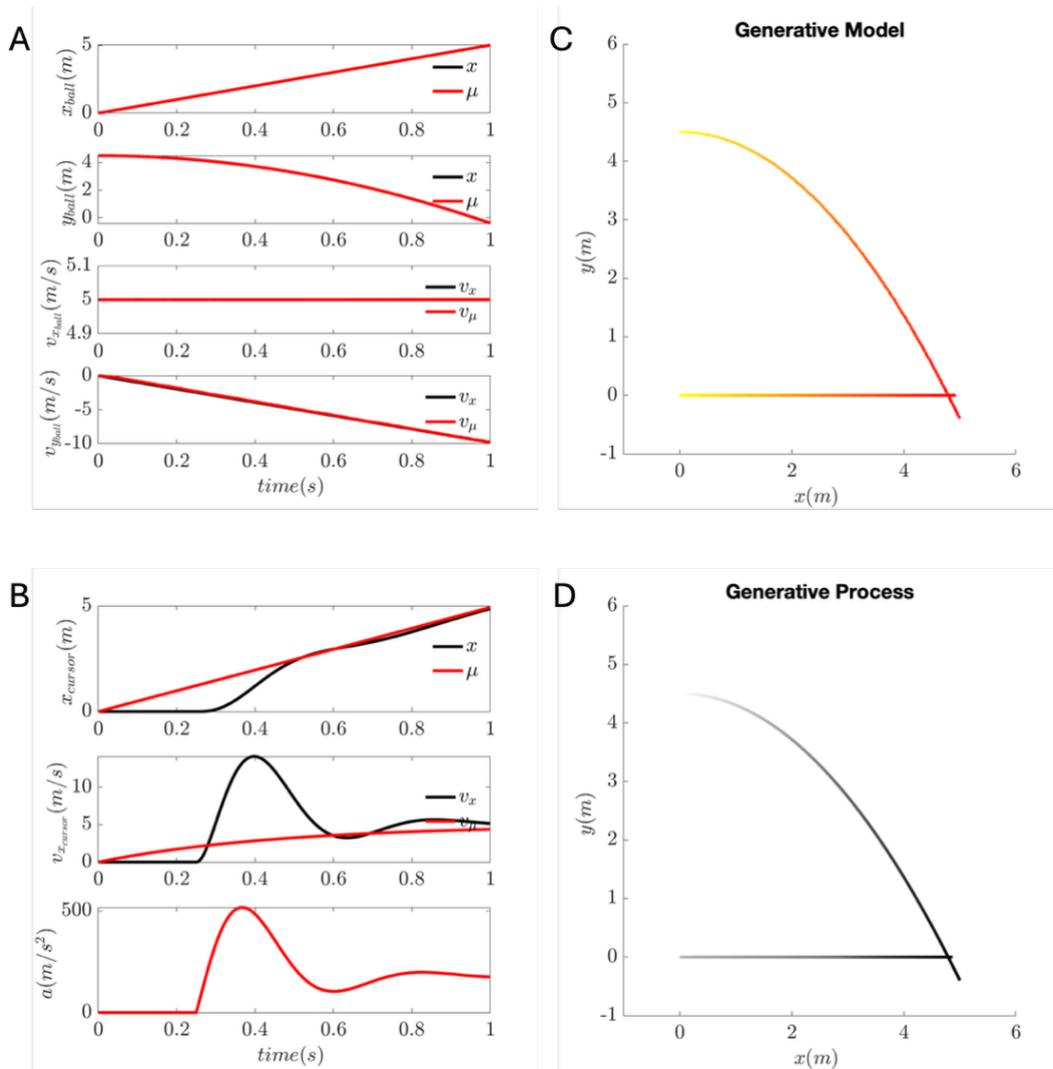

**Figure 2. Interception via *Next Location No Gravity* Strategy.** Temporal and spatial behavior of a moving ball intercepted by a horizontally moving cursor. Generative process trajectories are shown in black, while generative model (inferred) trajectories are shown in red. **(A)** *Temporal evolution of the ball state.* Real (black) and inferred (red) position ($x_{ball}$, $y_{ball}$) and velocity ($v_{x_{ball}}$, $v_{y_{ball}}$) of the ball are plotted over time for both horizontal and vertical components. **(B)** *Temporal dynamics of the cursor and control action.* The cursor's horizontal position ($x_{cursor}$) and velocity ($v_{x_{cursor}}$) are shown in black (real) and red (inferred). The bottom plot displays the magnitude of the control action $a$, represented here as acceleration. **(C)** *Inferred spatial trajectories of the ball and cursor.* A 2D projection of the inferred ball trajectory from (A), color-coded to indicate temporal progression (from yellow to red). The inferred cursor trajectory is also shown. **(D)** *Real spatial trajectories of the ball and cursor.* Actual trajectories of both ball and cursor are shown in the $xy$-plane. Color shading indicates temporal progression along the paths (from grey to black).

*Next Location With Gravity Strategy*

With gravity prior, the ball's estimated position and velocity remain highly accurate (Figure 3A), but the internal estimate of the cursor velocity is noticeably noisier (Figure 3B). Figures 3C and 3D show the x–y trajectories for the generative model and the actual process. As before, the action does not converge to zero, and the cursor moves smoothly toward the ball but does not achieve the goal to intercept it.

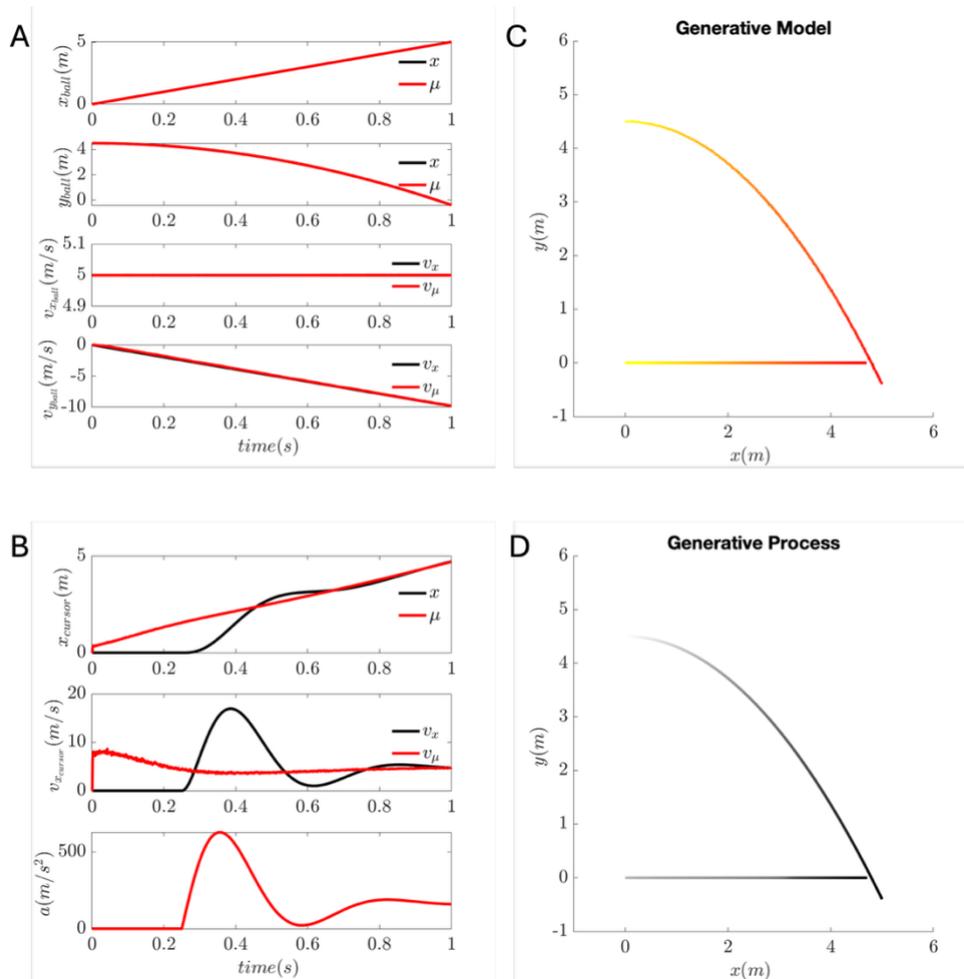

**Figure 3. Interception via *Next Location With Gravity* Strategy.** Temporal and spatial behavior of a moving ball intercepted by a horizontally moving cursor, using an *Next Location* generative model with gravitational acceleration. Generative process (real) trajectories are shown in black, and inferred trajectories from the generative model are shown in red. **(A)** *Temporal evolution of the ball state.* Real (black) and inferred (red) position $(x_b, y_b)$ and velocity $(V_{x_b}, V_{y_b})$ of the ball plotted over time, showing both horizontal and vertical components. **(B)** *Temporal dynamics of the cursor and control action.* Real (black) and inferred (red) horizontal position $(x_c)$ and velocity $(V_{x_c})$ of the cursor. The bottom plot depicts the control action $a$, represented as acceleration over time. **(C)** *Inferred spatial trajectories of the ball and cursor.* The 2-D inferred trajectories of ball and cursor are shown. Color from yellow to red indicates temporal evolution. **(D)** *Real spatial trajectories of the ball and cursor.* Actual paths of both ball and cursor in the $xy$-plane, with color shading indicating temporal progression (from grey to black).

## Interceptive Location No Gravity Strategy

This implementation predicts the ball's final interception point rather than its next-step position. The ball's position and velocity estimates (red) again match the true values (black) closely (Figure 4A), while the internal estimates of the cursor's position and velocity diverge from reality. These deviations actually steer the executed movement toward inferred cursor state which converge on its real values only near the end of the simulation (Figure 4B). Figures 4C and 4D display the generative model and the real process in the x–y plane. The action signal persists throughout the trial—never decaying to zero—indicating sustained control of the movement.

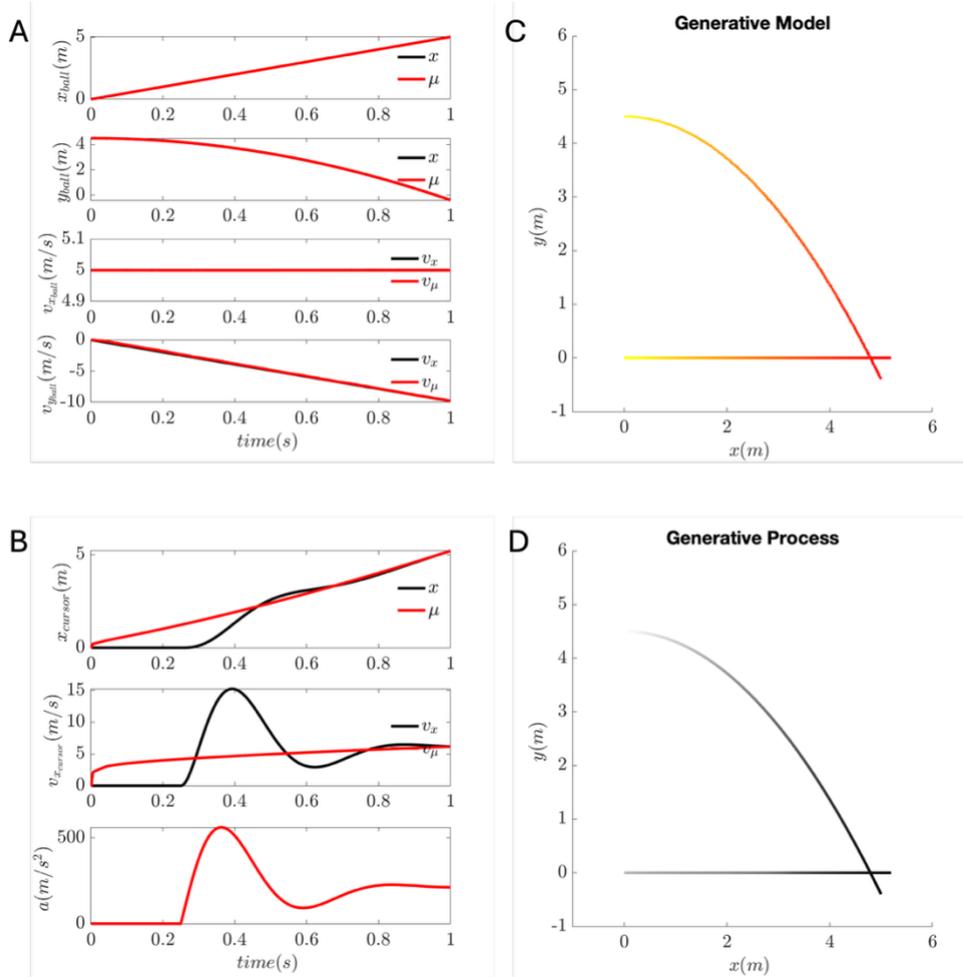

**Figure 4. Interception via *Interceptive Location No Gravity* Strategy.** Temporal and spatial behavior of a moving ball intercepted by a horizontally moving cursor, using an *Interceptive Location No Gravity* generative model. Generative process (real) trajectories are shown in black, and inferred trajectories from the generative model are shown in red. **(A)** *Temporal evolution of the ball state.* Real (black) and inferred (red) position ($x_b$, $y_b$) and velocity ($V_{x_b}$, $V_{y_b}$) of the ball plotted over time, covering both horizontal and vertical dynamics. **(B)** *Temporal dynamics of the cursor and control action.* Horizontal position ($x_c$) and velocity ($V_{x_c}$) of the cursor are shown in black (real) and red (inferred). The control action $a$, represented as acceleration, is shown in the bottom panel. **(C)** *Inferred spatial trajectories of the ball and cursor.* The 2D inferred trajectories of both ball and cursor are shown with time-progressive color coding (from yellow to red), reflecting how the *Interceptive Location No Gravity* estimation shapes the interception behavior. **(D)** *Real spatial trajectories of the ball and cursor.* The actual motion paths of the ball and cursor in the $xy$-plane, with shading indicating the progression through time (from grey to black).

## Interceptive Location With Gravity Strategy

When gravity is included, the ball's state continues to be estimated accurately (Figure 5A), but the cursor state behaves differently. The internal position estimate $\mu_{xc}$ shows pronounced high-frequency noise and rapidly advances toward the final position; the generative process follows this estimate with two oscillations (Figure 5B, first panel). The velocity estimate $\mu_{V_{x_c}}$ deviates from the true cursor velocity (Figure 5B, second panel) because the latter is largely driven by the action. The action signal $a$ starts with a large amplitude, then oscillates around zero as the simulation ends (bottom panel). Figures 5C and 5D illustrate the generative model and the real process in the x–y plane, revealing qualitatively similar overall trajectories.

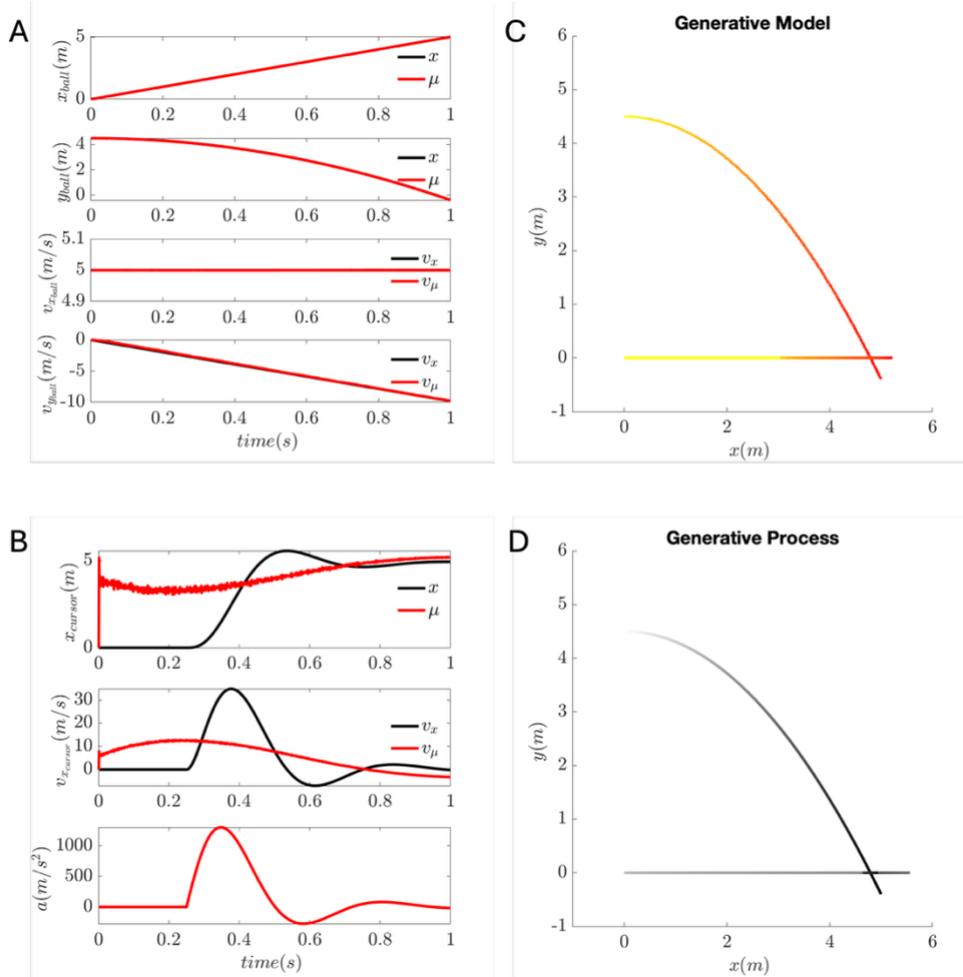

**Figure 5. Interception via *Interceptive Location* Strategy *With Gravity*.** Temporal and spatial behavior of a moving ball, subject to gravitational acceleration, intercepted by a horizontally moving cursor using a *Interceptive Location* generative model. Generative process (real) trajectories are shown in black, and inferred trajectories from the generative model are shown in red. **(A)** *Temporal evolution of the ball state.* Real (black) and inferred (red) position ($x_b, y_b$) and velocity ($V_{x_b}, V_{y_b}$) of the ball over time. The parabolic motion resulting from gravity is evident in the vertical components. **(B)** *Temporal dynamics of the cursor and control action.* Horizontal position ($x_c$) and velocity ($V_{x_c}$) of the cursor are shown for both the real (black) and inferred (red) trajectories. The bottom panel shows the control action $a$, represented as acceleration. **(C)** *Inferred spatial trajectories of the ball and cursor.* 2D inferred trajectories of both ball and cursor, with a time-progressive color scale (yellow to red). **(D)** *Real spatial trajectories of the ball and cursor.* Actual motion paths of the ball and cursor in the $xy$-plane. The cursor follows a horizontal trajectory to intercept the ball. Color shading indicates temporal progression (from grey to black).

*Comparison of Interceptive Strategies*

The four strategies were evaluated across various initial conditions (i.e., *vb0* and *xc0*) using their respective optimal parameter sets identified through a prior optimization procedure. To assess strategy's performance, four metrics were considered: *Action Magnitude* (i.e., the absolute value of the action at the end of the simulation), *Spatial Error*, *Temporal Error,* and *Inversions* (i.e., the number of direction changes in the cursor trajectory).
Action is the drive that aims at minimizing the error between sensory information and internal estimate. In this agent the action is defined in terms of the acceleration that makes the cursor moves, in other words, action set at zero means that the internal estimate has reached the expected value.

Estimating the *Next Location* of the cursor movement led to higher action magnitudes at the end of the simulation for both *With Gravity* and *No Gravity* strategies. However, including gravity significantly improved performance. This effect is illustrated in Figure 6, which shows results for different initial ball velocities and cursor starting positions. Action at the end of the simulation was lower for both the *Interceptive Location* strategies (F(1,783) = 259.43, p < 0.001) and for strategies that included gravity (F(1, 783) = 132.22, p < 0.001). Furthermore, a significant interaction in the ANOVA revealed that the *Interceptive Location* model with gravity had the lowest action magnitude (F(1, 783) = 162.84, p < 0.001).

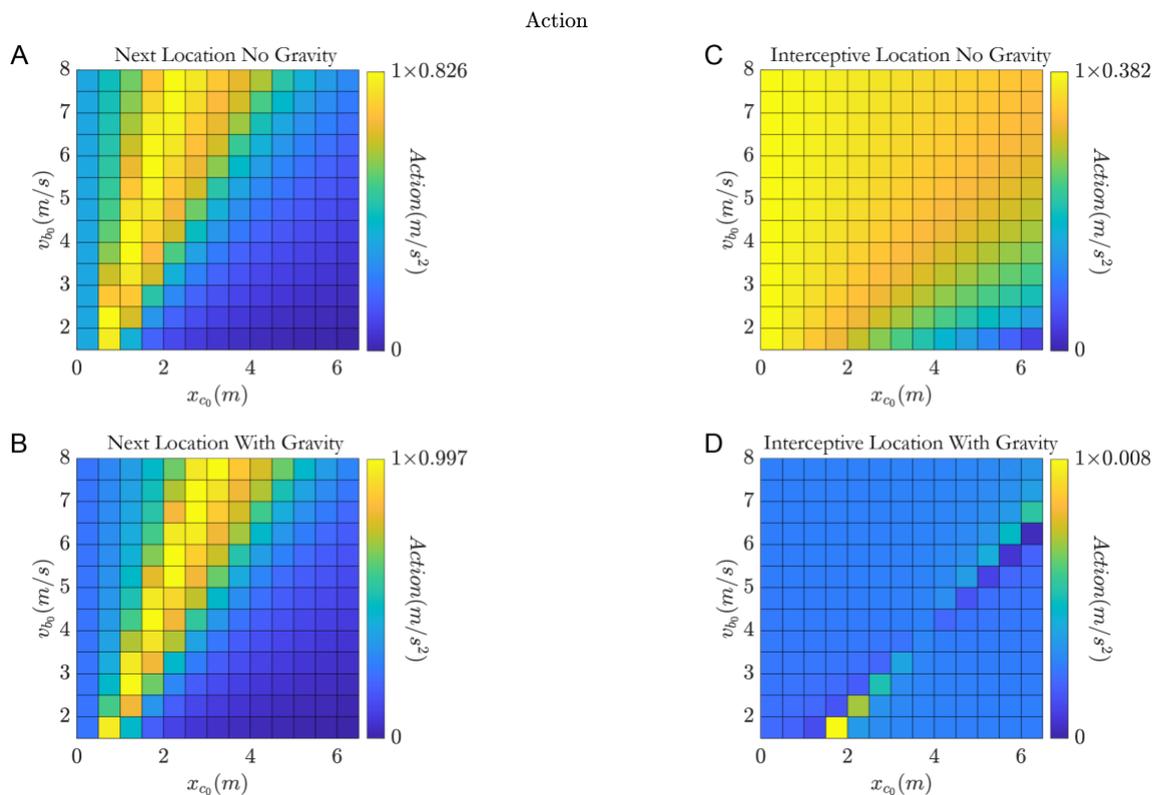

**Figure 6. Strategies comparison: Action Magnitude.** Comparison of the four strategies applied to the Action metric. Each subplot shows the error over the initial condition space defined by $x_{c0}$ and $v_{b0}$. **A-B**: *Next Location* strategies; **C-D**: *Interceptive Location* strategies. **A-C**: strategies without gravity (No Gravity); **B-D**: strategies with gravity (With Gravity). Color indicates normalized error magnitude, with warmer colors representing higher errors.

Spatial error was defined as the absolute difference between the positions of the ball and the cursor at the moment the ball reaches the horizontal line where the cursor moves (i.e. $y_{ball} = 0$). The spatial error is the indication of how close the cursor got to the ball during its motion. Results are shown in Figure 7. Overall, all strategies performed well on this metric, though the *Interceptive Location No Gravity* strategy performed worse—especially when the cursor started farther from the target (see Figure 9). A two-way ANOVA confirmed these findings, with both temporal horizon (F(1, 783) = 798.56, p < 0.001) and gravity (F(1, 783) = 544.25, p < 0.001) showing significant effects. The interaction was also significant (F(1, 783) = 531.35, p < 0.001), indicating that the *Interceptive Location No Gravity* model had larger spatial errors than the others.

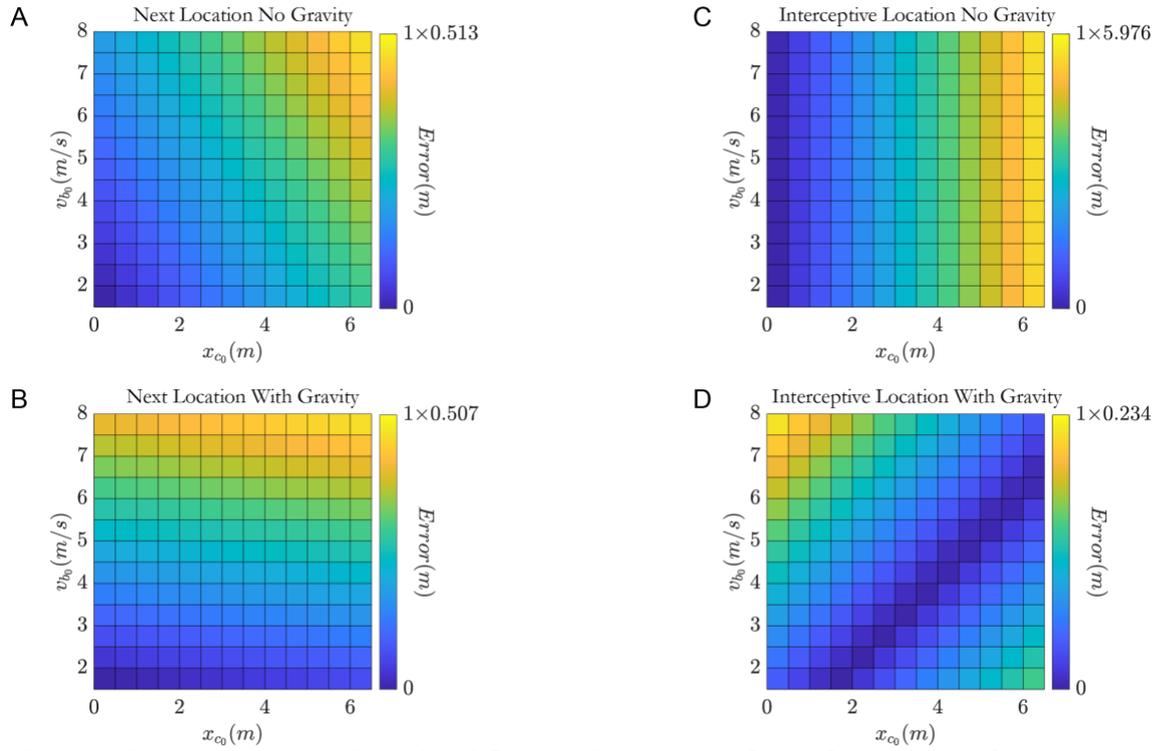

**Figure 7. Strategies Comparison: Spatial Error.** Comparison of spatial error across four strategies. Each subplot illustrates the spatial error as a function of initial conditions $x_{c0}$ and $v_{b0}$. **A-B**: *Next Location* strategies; **C-D**: *Interceptive Location* strategies. **A-C**: strategies without gravity (No Gravity); **B-D**: strategies with gravity (With Gravity). The color scale represents the magnitude bof spatial error, with brighter colors indicating greater deviation from the ideal position.

Temporal error, defined as the absolute time difference at the point of minimum distance between the ball and cursor trajectories, is shown in Figure 8. This metric indicates how far in time the cursor was when it reached closest to the ball. All strategies demonstrated small temporal errors across initial conditions, indicating that the cursor closely approached the ball. Statistical analysis revealed a significant interaction ($F(1, 783) = 20.85$, $p < 0.001$), suggesting the *Interceptive Location No Gravity* model differed from the others – being the strategy with the largest errors. Temporal error also varied significantly with both temporal horizons considered ($F(1, 783) = 7.01$, $p < 0.01$) and gravity prior ($F(1, 783) = 11.69$, $p < 0.001$).

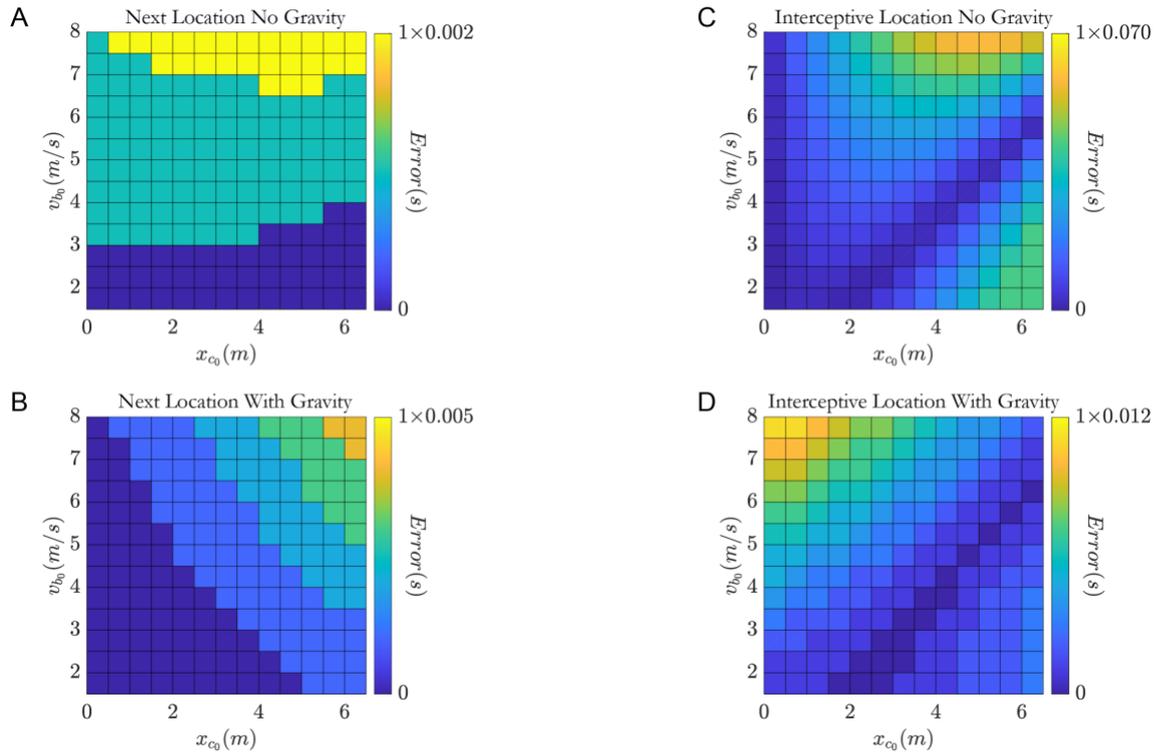

**Figure 8. Strategies Comparison: Temporal Error.** Comparison of temporal error across four strategies. Each subplot illustrates the temporal error as a function of initial conditions $x_{c0}$ and $v_{b0}$. **A-B**: *Next Location* strategies; **C-D**: *Interceptive Location* strategies. **A-C**: strategies without gravity (No Gravity); **B-D**: strategies with gravity (With Gravity). The color scale represents the magnitude of spatial error, with brighter colors indicating greater deviation from the ideal position.

Lastly, model performance was assessed by the number of direction changes in the cursor path. With this metric we aim at evaluating the smoothness of the cursor path, i.e. more inversions indicate a less smooth path. A clear difference was observed between *Next Location* and *Interceptive Location* strategies, with the latter showing a consistent number of reversals regardless of initial conditions (Figure 9). Indeed, the number of inversions were statistically significant according to two-way ANOVA for temporal horizon (F(1, 783) = 70.8, p <0.001) and not for gravity prior (F(1, 783) = 0.03, p = 0.85).

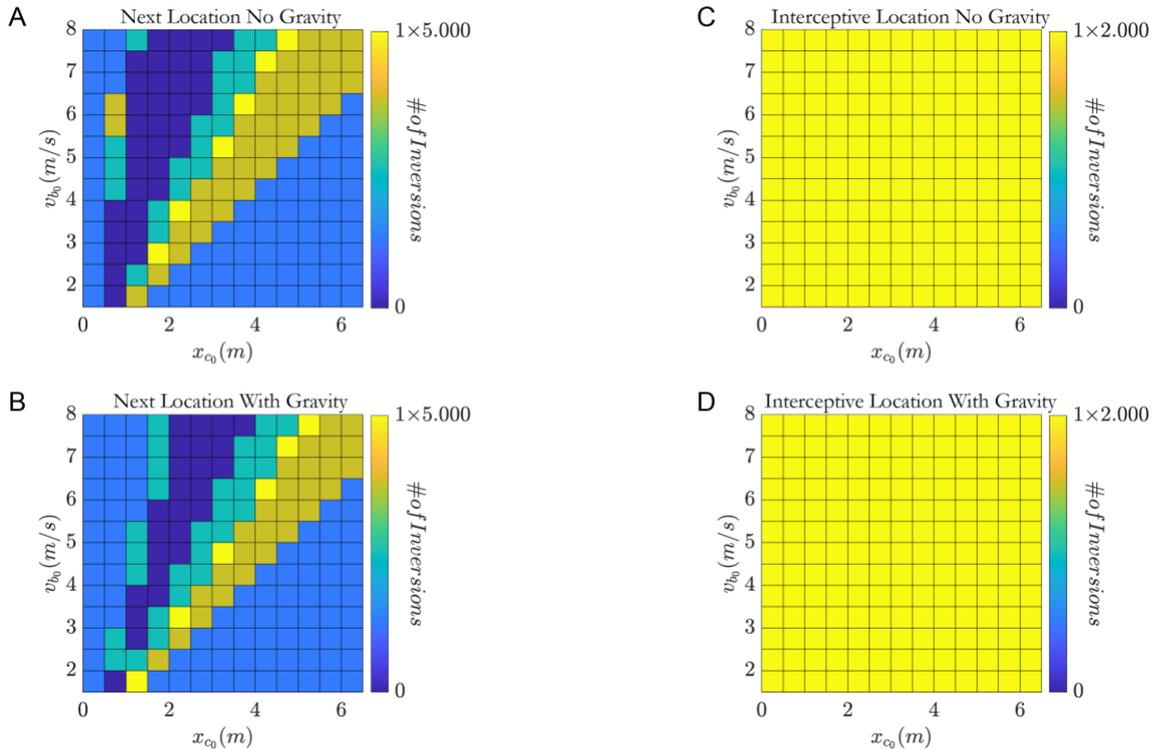

**Figure 9. Strategies Comparison: Inversions.** Comparison of the number of inversions across the four strategies. Each subplot illustrates the number of change in direction as a function of initial conditions $x_{c0}$ and $v_{b0}$. **A-B**: *Next Location* strategies; **C-D**: *Interceptive Location* strategies. **A-C**: strategies without gravity (No Gravity); **B-D**: strategies with gravity (With Gravity). The color scale represents the magnitude of spatial error, with brighter colors indicating greater deviation from the ideal position.

Figure 10 summarizes model comparisons by averaging all metrics across initial conditions. Across metrics, clear differences emerged between strategies. *Action Magnitude* (Figure 10A) was substantially lower for the *Interceptive Location* strategies, with the *Interceptive Location With Gravity* approach exhibiting the smallest final action, consistent with the agent settling into the estimated location. *Spatial Error* (Figure 10C) was generally low for all strategies, although the *Interceptive Location No Gravity* approach showed noticeably larger deviations. *Temporal Error* (Figure 10D) remained small across strategies, indicating tight temporal alignment between ball and cursor trajectories; however, the *Interceptive Location No Gravity* approach again showed the largest deviations. Finally, the number of inversions (Figure 10B) revealed a strong distinction in movement smoothness: *Interceptive Location* strategies showed consistent and relatively low numbers of direction changes, while *Next Location* strategies produced more reversals in cursor movement.

Together, these results confirm that incorporating gravity and aiming for the location of possible interception yields more accurate, temporally aligned, and smoother trajectories compared to strategies relying solely on *next location* estimation. Indeed, among all the strategies, the *Interceptive Location With Gravity* approach consistently outperformed the others across all four metrics.

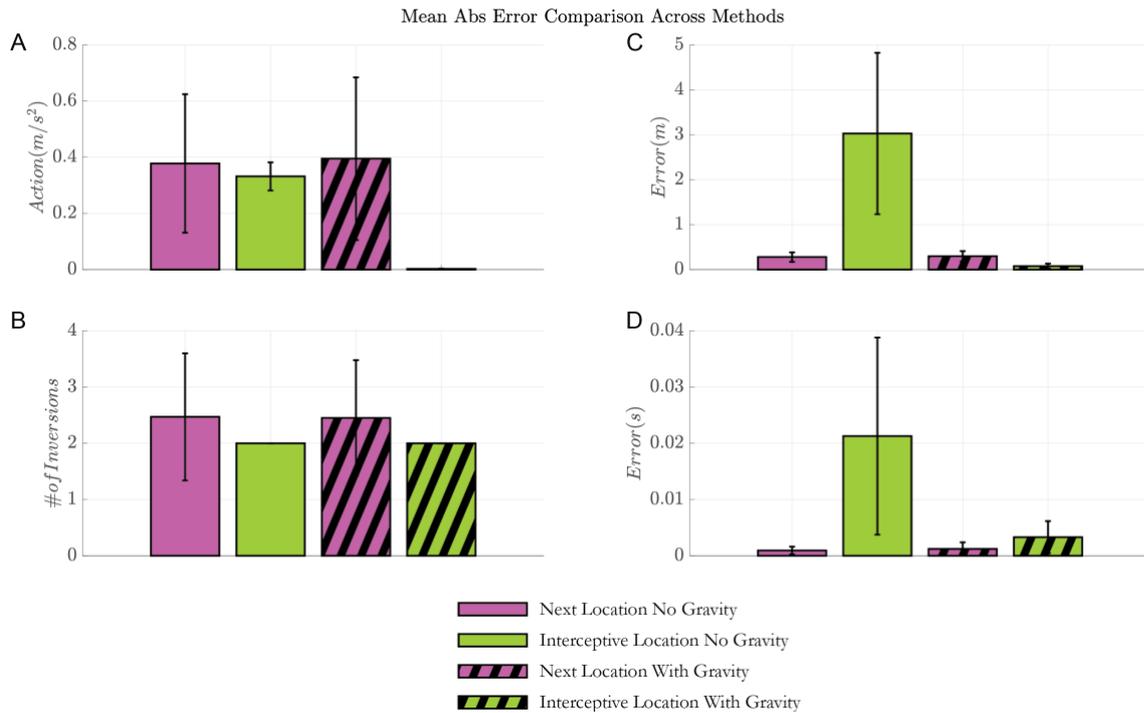

**Figure 10. Strategies Comparison: Summary.** Mean and standard deviation across initial conditions of the four performance metrics across strategies. Each subplot shows a different metric: Action (**A**), Inversions (**B**), Spatial Error (**C**), and Temporal Error (**D**). Bars represent the average value for each strategy pooling across all the initial conditions tested in the simulations. Error bars indicate the standard deviation. The four strategies evaluated are *Next Location* and *Interceptive Location* estimators, both with and without gravity modeling (*No Gravity* vs. *With Gravity*).

*Model robustness*

To examine how sensory uncertainty affects motor responses, we simulated the catching strategies under increasing levels of noise applied to the perceived ball trajectory (see Supplementary Materials). The effects of sensory noise on the cursor's trajectories is shown in Figure S1. Interceptive performance quantified through the temporal and spatial errors are shown in Figure S2A: temporal error remained small for moderate noise levels except for the *Interceptive Location* variant of the models, which resulted in larger errors as soon as noise got larger than the one used for the optimizations.

*Spatial error* (Figure S2B) followed a similar trend. Although for low noise, the cursor was minimally affected and intercepted the ball successfully, at higher noise levels, spatial deviations of the cursor path increased markedly, with the largest errors observed in the *Interceptive Location* variants.

Overall, the analysis demonstrates that the model is robust to moderate sensory uncertainty, but a proper tuning of the parameters might improve its behavior even for larger noise levels.

**Discussion**

The present study investigates predictive strategies for interceptive actions within the framework of Active Inference, comparing strategies that differed in (i) whether they included an internal representation of gravity and (ii) how far into the future they

predicted — either to the *Next Location* or the *Interceptive Location* of the moving target. Our simulations demonstrate that all four strategies were able to generate successful interceptive movements, but their performance differed systematically across metrics and environmental conditions. Notably, including gravity in the generative model improved spatial and temporal accuracy, while *Interceptive Location* prediction enhanced energetic efficiency and overall stability of control. Together, these results suggest that internal representations of environmental dynamics—particularly gravitational acceleration— play a key role in minimizing prediction error.

Our results support the view that interceptive actions rely on predictive control grounded in generative models that encode fundamental physical regularities. Within the Active Inference framework, the inclusion of gravity can be interpreted as embedding a structural prior about the causal laws governing the environment [26,28], also consistent with recent hierarchical Active Inference models in which action selection and motor control emerge from predictions encoded at multiple levels of a generative model [53–56]. From a neurophysiological perspective these priors may be learned through long-term sensorimotor experience and embedded in neural circuits, such as the vestibular system and cerebellum, that support the hypothesis of dynamic prediction [57–59].

Indeed, a well-established principle in motor control is that humans rely on internal models, also referred to as world models or primitives, to compensate for delays in the sensorimotor system. These models help predict the outcomes of actions and environmental states. This hypothesis aligns with a Bayesian framework, in which an internal model (prior) is refined through sensory input to estimate a posterior and generate motor commands [60,61].

If we accept this perspective, it follows that internal models are continuously learned and refined throughout life. A compelling case study is the internal model of gravity. Some studies have investigated whether newborns have expectations regarding object behavior [62–65]: from 2-months old they would expect object without support to fall.

Across all simulations of our model, the presence of a gravity prior substantially enhanced performance. Agents that included gravity in their internal model showed smaller spatial and temporal errors across different initial conditions. This supports the hypothesis that humans rely on an internal model of gravity that is automatically integrated into motor prediction [16,66]. Within the Active Inference framework, this internal model can be interpreted as a strong prior that constrains expectations about motion trajectories. By reducing uncertainty in vertical motion estimation, the gravity prior effectively biases inference toward more physically plausible trajectories, thus improving predictive control. These findings resonate with behavioral studies showing impaired interception performance in microgravity or altered-gravity contexts [19–21,67], and extend them to a formal computational setting.

An important contribution of this work is the comparison between *Next Location* and *Interceptive Location* strategies, which differ in the temporal horizon of their predictions. *Next Location* strategies continuously updates beliefs about the immediate next position of the target, while *Interceptive Location* strategies directly predicts the endpoint of the trajectory. Our results indicate that *Interceptive Location* strategies, particularly when coupled with gravity, outperformed *Next Location* strategies in accuracy and efficiency.

This suggests that longer prediction horizons can stabilize control by reducing corrections, effectively smoothing the sensory–motor loop. This finding is in line with a relevant set of studies which has shown that human subjects have been found to make anticipatory saccades to locations towards the bounce point when catching a ball [68–70].

A key advantage of the Active Inference formulation is its robustness to uncertainty. As shown in the supplementary material, our noise manipulation analyses revealed that model performance was stable across a range of sensory noise levels. This mirrors human performance, which typically remains accurate despite substantial visual noise and processing delays. Interestingly, *Next Location* strategies were slightly more resilient to extreme noise, possibly due to their shorter temporal dependency on past estimates.

A limitation of this study is that it focuses on a simplified one-dimensional interception task, which allows for precise evaluation of model performance but necessarily abstracts away from the complexity of real-world sensorimotor coordination. Extending this work to two- or three-dimensional trajectories, incorporating realistic sensory delays and proprioceptive feedback, and testing the model against human behavioral data would provide stronger ecological validity. Such extensions could include experiments on the kinematics of human behavior in more naturalistic tasks, paving the way for the investigation of richer motor features such as coarticulation [71,72]. Taking inspiration from robotics, future models could also include a more complex end effector, enabling the simulation not only of sensory feedback but also of the control of individual joints and their coordination [47,73–76]. Additionally, future implementations could incorporate learning mechanisms that allow the agent to adapt its internal model parameters over time, providing a more complete account of how internal models of environmental dynamics might emerge from experience [74,77,78].

In summary, this study demonstrates that Active Inference provides a powerful computational framework for modeling motor control, merging predictive strategies with online adjustments. Agents equipped with a gravity prior achieved more accurate and efficient interception, and those predicting the *Interceptive Location* of the target trajectory exhibited superior control stability. These findings support the hypothesis that structured priors about physical dynamics, such as gravity, can anticipate sensory consequences and guide rapid, adaptive action.

## Acknowledgements

This research received funding from the European Research Council under the Grant Agreement No. 820213 (ThinkAhead), the Italian National Recovery and Resilience Plan (NRRP), M4C2, funded by the European Union, NextGenerationEU (Project IR0000011, CUP B51E22000150006, "EBRAINS-Italy"; Project PE0000013, "FAIR"), and the Ministry of University and Research, PRIN PNRR P20224FESY and PRIN 20229Z7M8N. The funders had no role in study design, data collection and analysis, decision to publish, or preparation of the manuscript. We used a Generative AI model to correct typographical errors and edit language for clarity.

# Supplementary materials

To examine how sensory uncertainty affects motor responses, we simulated the catching strategies under increasing levels of noise applied to the perceived ball trajectory. While the generative process of the ball trajectory remained constant, higher noise levels progressively corrupted the perceived ball position and velocity signals, altering the generative model and consequently cursor responses.

Specifically, we varied the standard deviation (σ) of the additive white noise applied to the ball's observed state variables. Each noise level was simulated independently, while all other model parameters (e.g., dynamics, learning rates, and initial conditions) were kept constant. The tested noise levels corresponded to progressively larger variances (σ), covering the range from σ = 0.01 to σ = 0.5 (i.e. σ = [0.01,0.1,0.2,0.5]). When testing we kept the relation between position and velocity σ as set in the main simulation, i.e. having the noise variance for velocity one order of magnitude higher than the position noise. Temporal and spatial errors were used to evaluate model robustness to noise levels. Figure S1 illustrates the different levels of noise that have been tested.

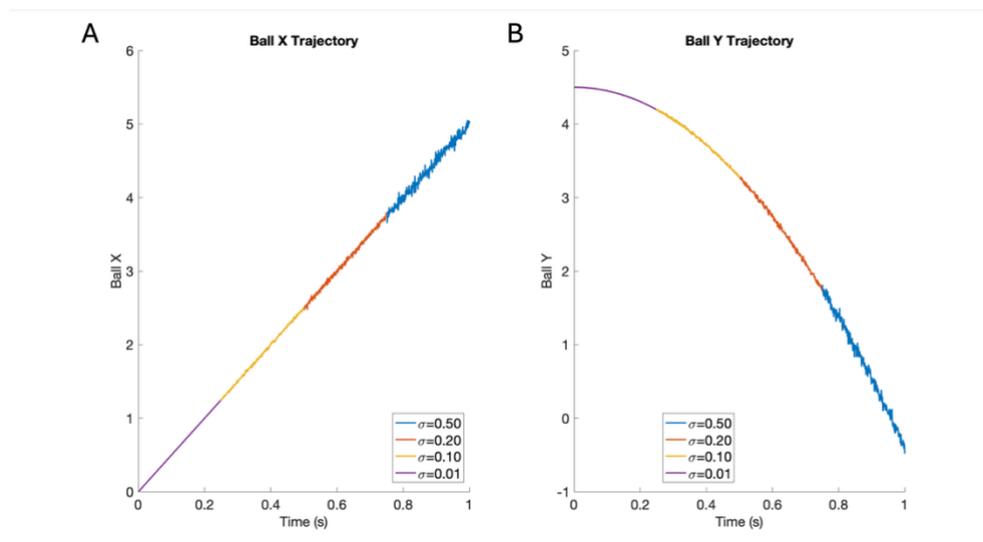

**Figure S1. Effect of sensory noise on perceived ball trajectories.** Horizontal (left) and vertical (right) components of the ball's observed trajectory for increasing levels of sensory noise (σ = [0.01,0.1,0.2,0.5]).

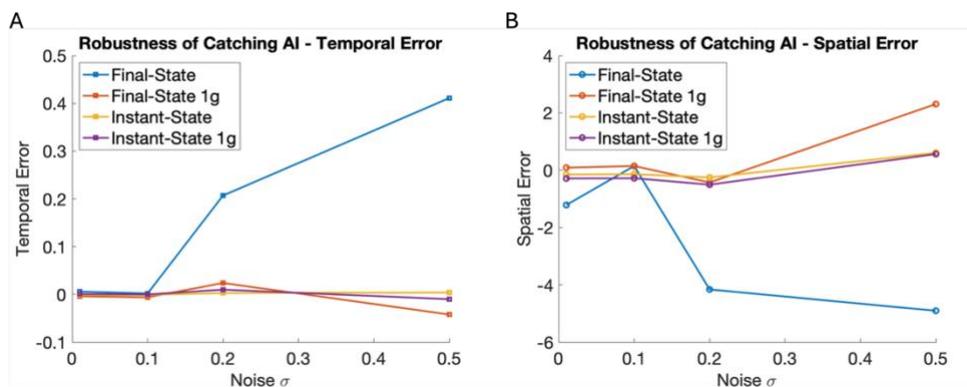

**Figure S2. Effects of sensory noise on catching performance.** (A) Temporal error as a function of sensory noise. Average temporal error between the ball and the cursor at the interception point across four model variants. (B) Spatial error as a function of sensory noise. Spatial error at interception, defined as the horizontal distance between the cursor and the ball when the ball reaches the target height.